\documentstyle[psfig]{mn}

\title{
Self-Similar Shocked Accretion of Collisional Gas with Radiative Cooling}

\author[M. G. Abadi, R. G. Bower and J. F. Navarro]
{Mario G. Abadi$^{1,2}$, Richard G. Bower$^2$ and Julio F. Navarro$^3$ \\
$^1$Observatorio Astron\'omico, Universidad Nacional de C\'ordoba, Laprida
854, 5000 C\'ordoba, Argentina\\
$^2$Department of Physics, University of Durham, South Road, DH1 3LE, Durham,
United Kingdom\\
$^3$
{\rm CIAR} Scholar and Sloan Fellow. Department of Physics and Astronomy, University
of Victoria, Victoria, BC, V8P 1A1, Canada}

\date{\today}

\begin{document}

\maketitle

\begin{abstract}

We describe similarity solutions that characterize the collapse of collisional
gas onto scalefree perturbations in an Einstein-de Sitter universe. We consider
the effects of radiative cooling and derive self-similar solutions under the
assumption that the cooling function is a power law of density and temperature,
$\Lambda(T,\rho) \propto \rho^{3/2} T$. We use these results to test the ability
of Smooth Particle Hydrodynamics (SPH) techniques to follow the collapse and
accretion of shocked, rapidly cooling gas in a cosmological context. Our SPH
code reproduces the analytical results very well in cases that include or
exclude radiative cooling.  No substantial deviations from the predicted central
mass accretion rates or from the temperature, density, and velocity profiles are
observed in well resolved regions inside the shock radius. This test problem
lends support to the reliability of SPH techniques to model the complex process
of galaxy formation.

\end{abstract}

\section{Introduction}

Structure forms in hierarchically clustering universes as primordial dark matter
density fluctuations are amplified by gravity and collapse in a constantly
evolving population of virialized dark matter halos. In this scenario galaxies
are envisioned to form as baryons follow the dark matter collapse, dissipate
their kinetic energy through shocks, and radiate it away as they settle (and
form stars) in centrifugally supported structures at the center of dark
halos. Galaxies evolve afterwards as a result of mergers between protogalaxies
and of further accretion of intergalactic gas (White \& Rees 1978, Navarro \&
White 1993, Cen \& Ostriker 1993, Navarro, Frenk \& White 1994, Evrard, Summers
\& Davis 1994, Katz, Weinberg \& Hernquist 1996, Bryan et al 1998, Couchman,
Thomas, \& Pearce 1995, Yepes et al 1997, Navarro \& Steinmetz 1997, Tissera,
Lambas \& Abadi 1997, Steinmetz \& Navarro 1998).

Gravity, pressure gradients, hydrodynamical shocks, and the ability of gas to
radiate are therefore physical processes of crucial importance during the
formation of galaxies in a cosmological context. Numerical experiments intended
to simulate galaxy formation must therefore capture accurately these essential
ingredients on the many different levels of the hierarchy that coexist at a
given time. Unfortunately, detailed analytic solutions are not known for
relevant analogues of the complex galaxy formation process and it has been
difficult to assess properly the accuracy and reliability of these codes (Frenk
et al 1999).

Previous studies have therefore focussed on the sensitivity and convergence of
the results regarding numerical parameters such as the size of the grid used in
Eulerian hydrodynamical methods (Cen 1992, Bryan et al 1998) or the number of
particles used in particle-based methods such as the Smooth Particle
Hydrodynamics (SPH, see Gingold \& Monaghan 1977, Benz 1990, Hernquist \& Katz
1989, Navarro \& White 1993, Summers 1993 for general introductions to
SPH). Convergence as resolution improves is a necessary condition for
simulations to be reliable, but is often not sufficient to ensure that the
results are accurate and robust. There is clearly a need for analytic solutions
that describe physical situations similar to the galaxy formation scenario
envisioned in cosmological models and that can be used to gauge the performance
of cosmological hydrodynamical codes.

Spherical infall is one relevant situation for which a detailed solution is
known. Bertschinger (1985) first computed the detailed behaviour of collisional
gas being accreted onto a point mass perturber in an Einstein-de Sitter
universe. Assuming that only gravity, pressure gradients, and hydrodynamical
shocks control the gas behaviour, Bertschinger exploited the scalefree nature of
all these processes to derive similarity solutions that offer a useful testbed
for hydrodynamical codes (Navarro \& White 1993, Summers 1993).  One crucial
ingredient of the galaxy formation process is, however, missing from these
tests: radiative cooling. This is because the cooling function is the result of
atomic processes that are not independent of scale and therefore similarity
solutions of the spherical infall problem are not generally available when
radiative energy losses are included. This is true even in the very simplified
case when radiation transfer, heat conduction, and magnetic effects are
neglected.

Self-similarity may still be recovered in situations that include radiative
cooling at the expense of placing restrictions on the temperature and density
dependence of the cooling function. For example, Bertschinger (1989) computed
the detailed self-similar evolution of cooling flows in isothermal potentials
under the assumption that the gas cooling function is a power law of density and
temperature, $\Lambda(\rho,T) \propto \rho^2 T^{\lambda}$. For $\lambda=1/2$,
this power law resembles the contribution from thermal bremsstrahlung to the
overall cooling function, so these results can be usefully applied to the hot,
diffuse X-ray emitting gas that fills the intracluster medium of rich galaxy
clusters. Unfortunately, these solutions are only valid for isolated systems
originally in hydrostatic equilibrium and therefore their applicability to
problems where continuous mass accretion play a significant role is limited.

A related approach has been recently described by Owen, Weinberg \& Villumsen
(1988), who derive a family of cooling functions that ensure self-similar
evolution in Einstein-de Sitter universes with power-law initial density
fluctuations. In this case, similarity is preserved by ensuring that the cooling
timescale of an object with characteristic clustering mass ($M_{\star}$) is a
fixed fraction of the Hubble time.

In this paper we follow a similar approach and derive similarity solutions for
the spherical infall problem that include energy losses due to radiative
cooling. Similarity is preserved by choosing a convenient power-law form of the
cooling function, and its solutions are compared with the results of direct
numerical simulations using a hydrodynamical SPH code. Because of the
restrictions placed on the cooling function, the applicability of the results to
realistic models of galaxy formation is not straightforward, but the solutions
are very useful as tests of hydrodynamical codes under physical conditions that
combine the major ingredients of galaxy formation models: gravitational
collapse, pressure gradients, energy dissipation through shocks, and radiative
energy losses, {\it in a proper cosmological context.}

We derive the similarity solutions in \S2 and compare them with numerical
simulations in \S3. Section 4 discusses the results and \S5 summarizes our main
conclusions.

\begin{figure*}
\centerline{\psfig{file=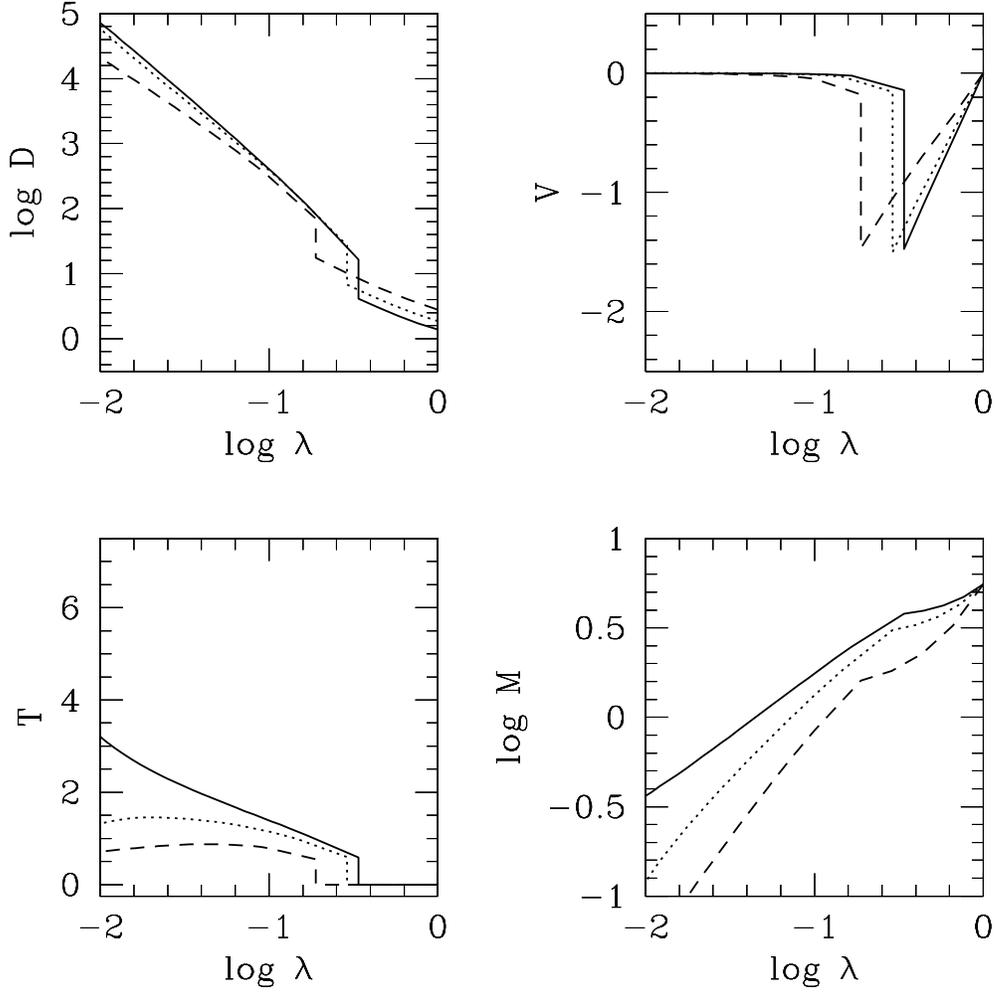,width=14cm}}
\caption{
Similarity solutions for different choices of the shape of the initial density
perturbation. The three curves show the solutions for $\epsilon=1$ (solid
lines), $2/3$ (dotted lines), and $1/3$ (dashed lines), respectively. Symbols
are as defined in eqs.(3). These analytic solutions neglect radiative energy
losses.}
\label{fig:nocool}
\end{figure*}

\begin{figure*}
\centerline{\psfig{file=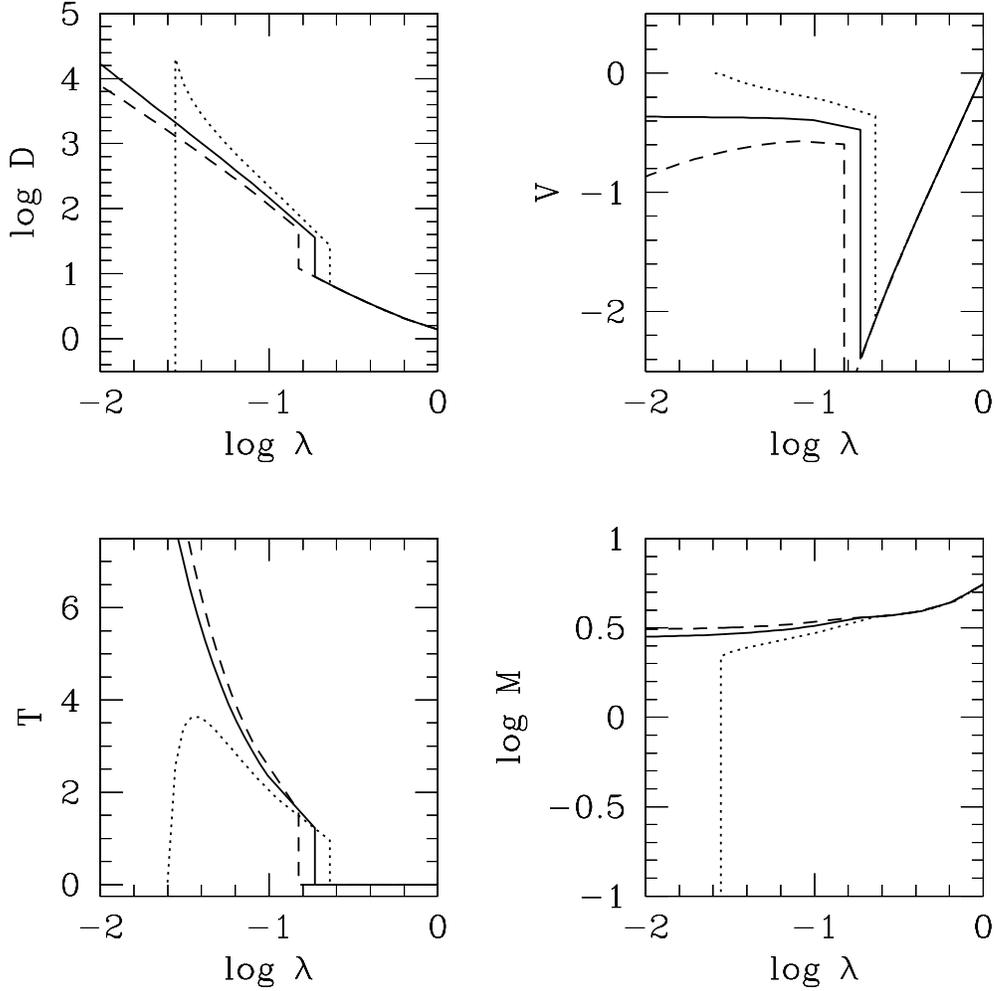,width=14.0cm}}
\caption{Examples of the three types of similarity solutions found when
radiative cooling is included, computed for the case $\epsilon=1$ and
$K_0=0.1$. The stagnation, adiabatic, and eigensolution correspond to the dotted
($\lambda_s=0.23$), dashed ($\lambda_s=0.15$), and solid line
($\lambda_s=0.1869$), respectively. The eigensolution represents a limiting case
of the two other kinds, when the stagnation point approaches the center and the
flow extends all the way to $\lambda=0$. }
\label{fig:lambdas}
\end{figure*}

\begin{figure*}
\centerline{\psfig{file=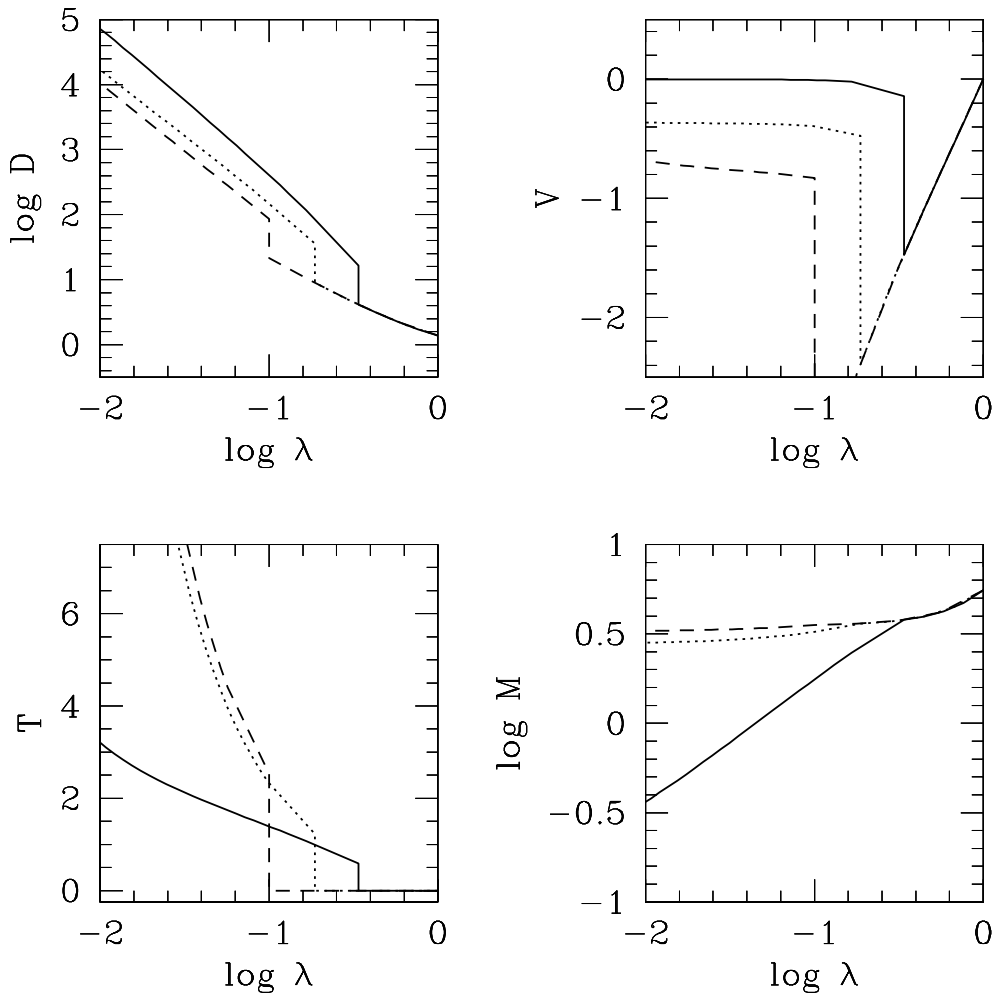,width=14cm}}
\caption{The density, velocity, temperature, and enclosed mass profiles showing the 
effect of increasing the relative importance of cooling. All curves correspond
to the ``eigensolution'' for $\epsilon=1$. Shock radii are given in Table 1. The
solid, dotted, and dashed curves correspond to $K_0 = 0.0$, $0.1$ and $0.3$,
respectively.}
\label{fig:k0plot}
\end{figure*}

\begin{figure*}
\centerline{\psfig{file=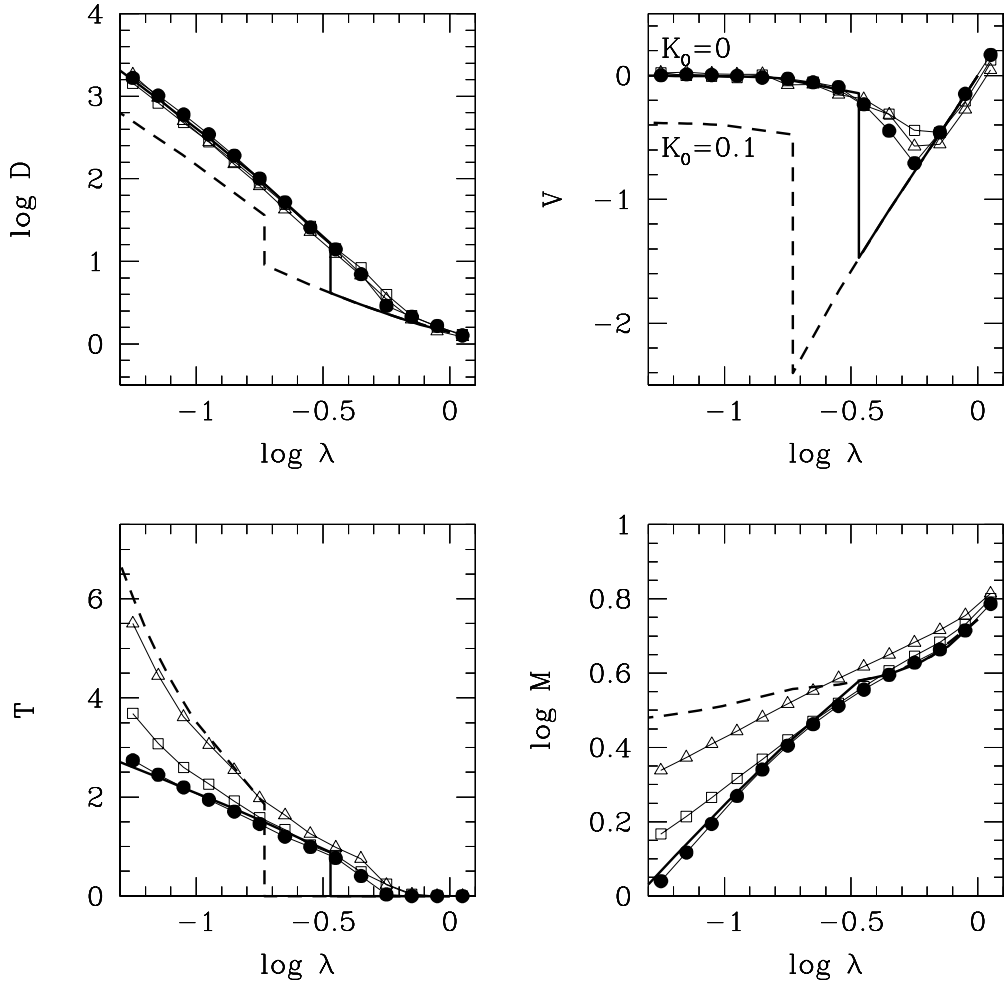,width=14cm}}
\caption{
Comparison between similarity solutions and the results of SPH simulations. The
solution without cooling is represented by the solid line. We also show, for
comparison, the solution including radiative energy losses
($K_0=0.1$). Different symbols correspond to the SPH simulation at different
times. Open triangles, squares and filled circles correspond to times when $\sim
7$, $15$, and $20\%$ of the initial mass lies inside the shock radius. The
results of the simulations are seen to converge to the analytic solution at
later times, as more particles pass through the shock and the effects of
numerical resolution become less important.}
\label{fig:nocool}
\end{figure*}

\begin{figure*}
\centerline{\psfig{file=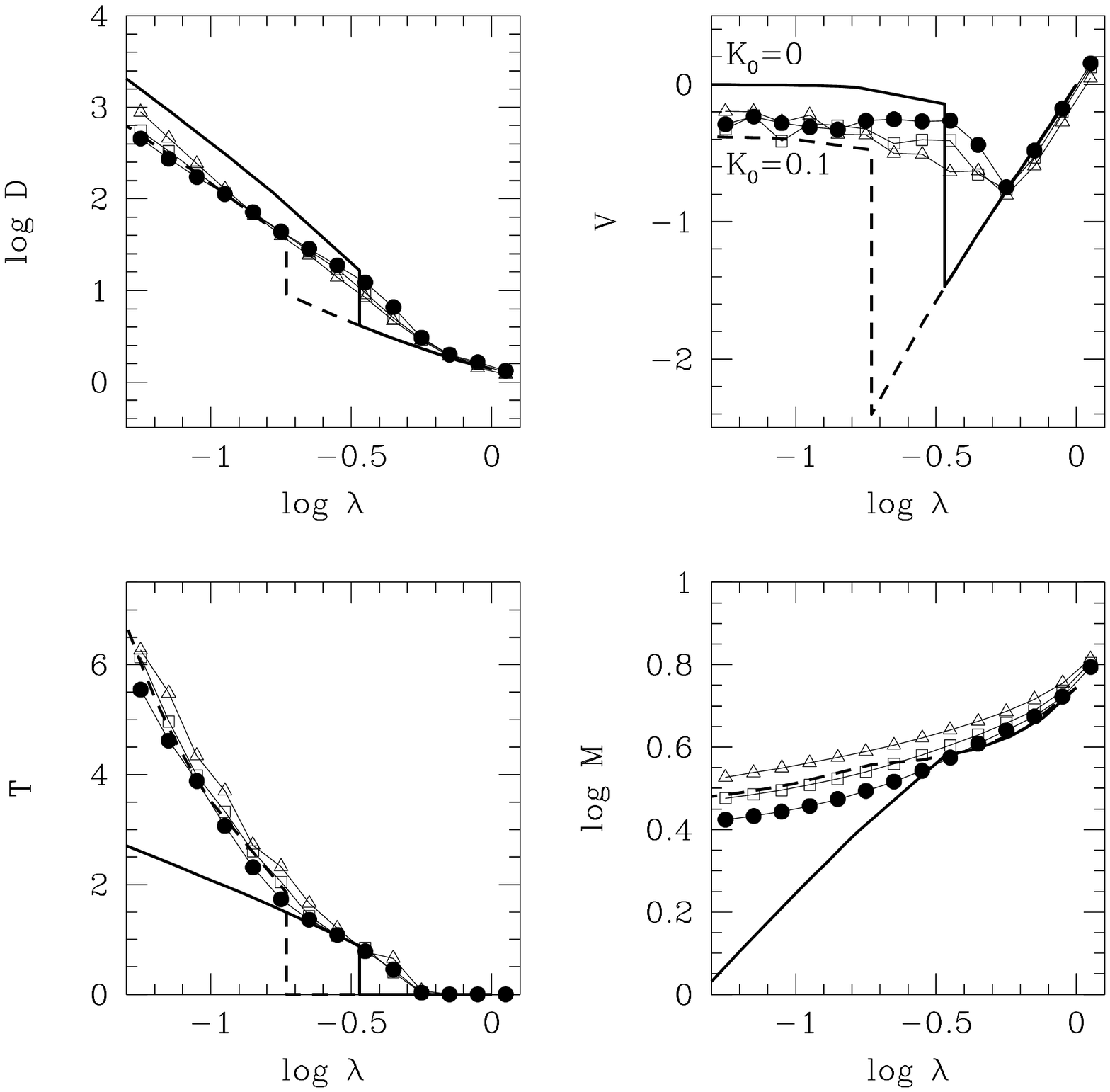,width=14cm}}
\caption{
As in Figure 4. The SPH simulation results shown now include cooling,
$K_0=0.1$. The times shown are analogous to those chosen in Figure 4. The SPH
simulation reproduces the analytical results very well inside the shock
radius. }
\label{fig:cool}
\end{figure*}

\section{Accretion of Collisional Gas onto Scale Free Perturbations}

\subsection{Similarity solutions neglecting radiative cooling}

The evolution of a density perturbation in an Einstein-de Sitter universe is
expected to be self-similar if the initial perturbation contains no physical
scales. Following Fillmore \& Goldreich (1984) we characterize the initial
perturbation at some initial time $t_i$ by the excess mass within a radial shell
of (initial) radius $r_i$,
$$
{\delta M_i \over M_i} = \left({M_i \over M_0}\right)^{-\epsilon}, \eqno(1)
$$
where $M_i=(4/3)\pi \rho_{H} r_i^3$ is the unperturbed mass within $r_i$,
$\rho_{H}=3H_i^2/8\pi G$ is the critical density for closure, $H_i$ is
Hubble's constant at $t=t_i$, $M_0$ is some reference mass, and
$\epsilon>0$. Because the mass excess is positive each radial shell is bound to
the center and collapses after reaching a (maximum) turnaround radius,
$r_{ta}$. The turnaround radius increases with time as
$$ 
r_{ta} \propto t^\xi, \eqno(2)
$$
where $\xi=(2/3)(1+1/3\epsilon)$. The mass inside the turnaround radius then
grows as $M_{ta}=M(r<r_{ta}) \propto t^{2/3\epsilon} \propto (1+z)^{-1/\epsilon}$,
where $z$ is the usual definition of redshift.

This scaling can be compared with that of the characteristic clustering mass in
a scalefree hierarchical clustering universe where the power spectrum of initial
density fluctuations is $P(k) \propto k^{n}$: $M_{\star}(z) \propto
(1+z)^{-6/(n+3)}$. Perturbations characterized by a given value of $\epsilon$
therefore accrete mass at the same rate as a ``typical'' mass concentration in a
scalefree universe with $n=3(2\epsilon-1)$. A central point mass perturbation
corresponds to $\epsilon=1$, with $r_{ta}\propto t^{8/9}$ and $M_{ta}\propto
t^{2/3} \propto (1+z)^{-1}$. This is the case considered by Bertschinger (1985).

Assuming that at $t=t_i$ the initial velocity field is pure unperturbed Hubble
flow, $v=H_i r=(2/3)r/t_i$, there are no further scales in the problem once the
magnitude of initial density perturbation has been specified, and the time
evolution of the system must approach self-similarity after a short initial
transient. This implies that a unique solution, expressed in properly scaled
variables, describes the structure of system at all times $t \gg t_i$. It is
convenient to express this solution in nondimensional form and, following
Bertschinger (1985), we define dimensionless radii, velocities, densities,
pressures, masses, and temperatures as follows,
$$
\lambda(r,t)={r\over r_{ta}} \eqno(3.1)
$$
$$
v(r,t)={r_{ta}\over t} V(\lambda) \eqno(3.2)
$$
$$
\rho(r,t)=\rho_H(t) D(\lambda) \eqno(3.3)
$$
$$
p(r,t)=\rho_H(t) \left({r_{ta} \over t}\right)^2 P(\lambda) \eqno(3.4)
$$
$$
m(r,t)={4\pi \over 3} \rho_H r_{ta}^3 M(\lambda) \eqno(3.5)
$$
$$
T(\lambda)={P(\lambda) \over (\gamma-1) D(\lambda)}, \eqno(3.6) 
$$
where $\gamma=5/3$ is the usual ratio of specific heats.

The equations describing the motion of a collisional fluid with spherical
symmetry can be expressed in terms of these dimensionless variables and are
given by (see Bertschinger 1985 for details),
$$
(V-\xi\lambda)D' + DV' + {2DV\over\lambda} -2D = 0 \eqno(4.1)
$$
$$
(V-\xi\lambda)V'-(1-\xi)V=-{P'\over D} - {2\over9}{M\over\lambda^2} \eqno(4.2)
$$
$$
(V-\xi\lambda)\left({P'\over P}-\gamma{D'\over D}\right) = 2(2-\xi) - 2\gamma
\eqno(4.3)
$$
$$
M' = 3\lambda^2 D\eqno(4.4)
$$
Here primes refer to differentiation relative to $\lambda$.  Eqs.(4) are,
respectively, the continuity, Euler, adiabatic, and mass equations and are valid
for a pressurized fluid flow neglecting radiation, heat conduction, and
deviations from spherical symmetry.

Pressurization of each radial shell of fluid occurs soon after turnaround as the
collapsing shell encounters previously collapsed ones. Because of similarity
constraints, the radius at which the shock occurs must be a constant fraction of
the turnaround radius, $\lambda=\lambda_s$.  Outside $\lambda_s$ the evolution
of the gas is identical to the turnaround and collapse of a pressureless shell
of material. A full solution of eqs.(4) can be found by locating the radius of
the shock and applying Hugoniot shock jump conditions to the exterior
pressureless infall values. A parametric form of the preshock cold accretion
flow is given by,
$$
\lambda = \sin^2(\theta/2) \,\left(\theta-\sin\theta\over\pi\right)^{-\xi}
\eqno(5.1)
$$
$$
V(\lambda) = \lambda {\sin\theta(\theta-\sin\theta)\over(1-\cos\theta)^2}
\eqno(5.2)
$$
$$
D(\lambda) ={9\over2}{(\theta-\sin\theta)^2\over(1-\cos\theta)^3(1+3\epsilon\chi)} \eqno(5.3)
$$
$$
M(\lambda) = \lambda^3{9\over2}{(\theta-\sin\theta)^2\over(1-\cos\theta)^3}
\eqno(5.4)      
$$
with $\chi = 1 - (3/2)(V(\lambda)/\lambda)$.

The shock location depends also on the central boundary conditions, which we
take to be that the velocity and mass must vanish, i.e. $V=M=0$ at
$\lambda=0$. Identifying the values of the variables inside (outside) the shock
by the subscript 2 (1), we have, at $\lambda=\lambda_s$,
$$
{(V_2-\xi\lambda_s)\over(V_1-\xi\lambda_s)}={\gamma-1 \over \gamma+1}
\eqno(6.1)
$$
$$
D_2=\left(\gamma-1 \over \gamma+1\right) D_1  \eqno(6.2)
$$
$$
P_2={2\over\gamma+1}D_1(V_1-\xi\lambda_s)^2 \eqno(6.3)
$$
$$
M_2=M_1. \eqno(6.4)
$$

Figure 1 shows the resulting density, velocity, temperature and entropy profiles
for various values of the initial perturbation parameter $\epsilon$. As
$\epsilon$ decreases from unity (the value corresponding to a point mass
perturbation, see solid line) to $1/3$ (dashed line), the shock moves inwards,
the inner density profile becomes shallower and the temperature profile becomes
approximately isothermal. We shall see next how these results are altered by the
inclusion of radiative cooling effects.

\subsection{Similarity solutions including radiative energy losses}

\subsubsection{The self-similar cooling function}

The results discussed in the previous subsection are only applicable in the
limiting case when energy losses due to radiative cooling are neglected. As
discussed in \S1, the cooling function of a plasma with realistic cosmic
abundances has a complex dependence on temperature and imposes dimensional
physical scales on the problem that violate the conditions required for the
existence of self-similar solutions.

Similarity solutions may exist only when the cooling processes introduce no
further scales in the problem. This condition can be satisfied by choosing an
appropriate cooling function so that the overall cooling efficiency is
independent of time. This can be ensured by demanding, for example, that the
cooling radius (i.e. the radius at which the cooling time equals the age of the
universe) be a fixed fraction of the turnaround radius of the system.
Equivalently, one may require that, at some fixed fraction of the turnaround
radius, the ratio between the local cooling timescale and the age of the
universe be constant and independent of time.

The cooling time is given by
$$
t_{cool}={u\over du/dt}={u \rho \over \Lambda(\rho,T)} \propto {u \rho_H \over
\Lambda(\rho,T)}, \eqno(7)
$$
where $u\propto T$ is the specific thermal energy of the gas, and the
proportionality in eq.(7) is valid at a fixed value of $\lambda$.  The condition
$$
t_{cool}/t_H=(6\pi G\rho_H)^{1/2} t_{cool}=K_0^{-1}={\rm constant} \eqno(8)
$$
is thus satisfied if 
$$
\Lambda(\rho,T) \propto \rho^{3/2} u \propto \rho^{3/2} T. \eqno(9)
$$
This condition is independent of $\epsilon$ and implies that the solution will
be self similar regardless of the time dependence of the turnaround radius.

Our similarity solutions thus require a weaker dependence on density and a
stronger dependence on temperature than expected from thermal bremsstrahlung
emission, $\Lambda_{bremss} \propto \rho^2 T^{1/2}$. We note, however, that
eq.(9) is not the only cooling function that would lead to self-similar
evolution. In particular, since the characteristic ``virial temperature''
($T_{vir} \propto GM_{ta}/r_{ta}$) of the system is related to its mean density
by the growth rate of the turnaround radius, it is possible to retain the
$\rho^2$ dependence characteristic of realistic cooling functions and adjust
only the temperature exponent to preserve similarity. The price one pays is that
in this case the temperature exponent of the self-similar cooling function
depends on $\epsilon$. More explicitly, $\Lambda(\rho,T) \propto \rho^2
T^{\beta}$, with 
$\beta=1-(9/2)(\epsilon/(3\epsilon-2))$ 
(Owen et al 1998). In this case, the relative velocities of the cooling radius
and the shock radius are equal. We emphasize that our choice of self-similar
cooling function (eq.9) is independent of $\epsilon$ and does not rely on tuning
the two velocities to agree with each other.

\subsubsection{The similarity solutions}

Once the appropriate form of the cooling function has been chosen, the behaviour
of the gas can be computed using eqs.(4) after modifying the entropy
conservation equation (4.3) to allow for energy losses. In dimensionless form,
the modified eq.(4.3) now reads
$$
(V-\xi\lambda)\left({P'\over P}-\gamma{D'\over D}\right) = 2(2-\xi) - 2\gamma
-K_0 D^{1/2}. \eqno(10)
$$
At any radius inside the shock, and at all times, the ratio between the local
dynamical time ($3\pi/16G\rho)^{1/2}$ and the cooling time equals $\pi
\sqrt{18/16} K_0$.

Equations 4.1, 4.2, 4.4, and 10 can now be solved to describe the post-shock
flow once adequate boundary conditions at the center are imposed. As discussed
by Bertschinger (1989), three different kinds of solutions can be identified
according to the limiting behaviour of the solution near the center. The first
type is a solution where the flow stagnates at some finite radius. Infalling gas
settles onto this surface, where the density formally diverges. In order to obey
self-similarity the surface must move outwards at the same rate as the
turnaround radius. This kind of solution thus requires a piston to move the
surface outwards and is therefore of little physical applicability.

The second type of inner solution corresponds to a flow that extends all the way
to $\lambda=0$ so that the local flow time, $t_{flow}=r/v$, near the center
becomes much shorter than the cooling time. These solutions are referred to as
``adiabatic'' solutions, because cooling is unimportant for small $\lambda$. The
mass accretion rate near the center approaches a constant and the accretion
speed diverges near the center.

The third kind of solution, sometimes called the ``eigensolution'', is the
limiting adiabatic solution with minimum central mass accretion rate, or,
equivalently, the limit of the family of stagnating solutions as the stagnation
radius tends to zero.

Each of these solutions is characterized by different values of the shock
radius, $\lambda_s$. Values of $\lambda_s$ similar to those obtained neglecting
cooling correspond to solutions with stagnation points. As the shock radius
moves inwards the stagnation point moves closer to the center and the solution
transitions through the eigensolution to the adiabatic case.

Figure 2 shows examples of these three different solutions for the case
$\epsilon=1$, $K_0=0.1$. The solution with $\lambda_s=0.23$ (dotted line) has a
stagnation point at $\lambda_0 \sim 0.026$ where the density diverges and the
velocity becomes zero at the surface. There is no mass inside this radius, and
the surface is pushed out by a piston to preserve similarity. As the shock
radius is reduced to $\lambda_s=0.15$ the flow extends all the way to the center
and the infall velocity diverges there. The eigensolution corresponds to
$\lambda_s \approx 0.1869$. In this case the central velocity remains finite at
the origin and the flow extends all the way to the center. In practice, we find
this solution numerically by letting the stagnation point, $\lambda_0$, approach
zero.

\begin{table}
\begin{center}
\begin{tabular}{l|l|l|l}
\hline
  $\epsilon$ &    $\xi$ & $K_0$ & $\lambda_s$\\
\hline
        1 &     $8/9$ & 0.0 &   0.3389\\     
        1 &     $8/9$ & 0.1 &   0.1858\\    
        1 &     $8/9$ & 0.3 &   0.0939\\    
        \\
        2/3 &     1 &     0.0 &   0.2899\\
        2/3 &     1 &     0.1 &   0.1551\\
        2/3 &     1 &     0.3 &   0.0733\\
        \\
        1/3 &     $4/3$ & 0.0 &   0.1889\\
        1/3 &     $4/3$ & 0.1 &   0.0996\\
        1/3 &     $4/3$ & 0.3 &   0.0422\\
\hline
\end{tabular}
\caption{The dimensionless shock radius of the eigensolution for various choices
of the shape parameter of the initial density perturbation, $\epsilon$, and of
the dimensionless cooling coefficient, $K_0$. The time exponent of the
turnaround radius, $\xi$, is also listed for each case.}
\label{tab:solns}
\end{center}
\end{table}

Noting that the solutions have approximately constant infall velocity in the
inner regions, it is possible to determine the asymptotic slopes of the density
and pressure profiles. As the velocity tends to a constant,
$D(\lambda)\rightarrow D_0 \lambda^{-2}$ and $P(\lambda)\rightarrow
P_0\lambda^{-3}$, where the normalising constants depend on the cooling
parameter $K_0$.  The asymptotic velocity is given by $- K_0 D_0^{1/2} /
(2\gamma-3)$.

Figure 3 shows how the eigensolutions vary as a function of the cooling
efficiency parameter $K_0$. As the importance of cooling increases the pressure
support inside the shock decreases and the shock radius moves inwards. Perhaps
counterintuitively, as cooling becomes more important the temperature inside the
shock radius {\it increases} and the density {\it decreases}. This is because
low entropy gas is ``lost'' to the central mass and, at fixed radius, low
entropy gas is replaced by higher entropy gas that moves in from outside. The
``cooling flow'' thus results in a net {\it increase} in gas entropy at a given
radius. Figure 3 illustrates that cooling has a substantial effect on the
structure of the system, and suggests that similarity solutions with cooling may
provide a stringent test of the capabilities and accuracy of hydrodynamical
codes. We pursue this issue next.

\section{Comparison with SPH simulations}

As discussed in \S1, the solutions derived in the previous section can be
fruitfully confronted with the results of cosmological hydrodynamical
codes. This comparison is all the more interesting because the test case we
discuss in the previous section captures many of the salient features of the
galaxy formation process: gravitational collapse, pressurization through shocks,
radiative energy losses, cooling flows. Furthermore, because the solutions are
self similar in time, a single simulation can be examined at different times and
convergence can be directly assessed. This is important because the importance
of numerical resolution varies with time within a single simulation. For
example, the number of particles within the shock radius increases with time,
and the ratio between the smallest resolved radius and the turnaround radius
decreases with time. Because the solution is unique, analyzing the deviations
between analytic solution and numerical experiment at different times provides
invaluable insight into the role of numerical limitations and their consequence
on the subsequent evolution of the system.

We use the Smooth Particle Hydrodynamics code described by Navarro \& White
(1993), where details about the numerical procedure should be consulted. The
initial setup is also similar to that described by these authors. We simulate a
spherical region of an Einstein-de Sitter universe by laying down $24,257$
particles homogeneously inside a sphere of radius $R$. Each particle is given an
initial velocity consistent with unperturbed Hubble flow, and an external
potential is added to mimic a point mass perturbation of mass equal to $5\%$ of
the total mass of the sphere. The external potential is ``softened'' inside a
fixed radius $R_p=0.1R$ in order to prevent divergences, but is fully Keplerian
outside $R_p$. This corresponds to the case $\epsilon=1$ in eq.(1).

The gravitational softening of each particle is also chosen to be equal to
$R_p$. All particles have initially the same temperature, chosen to be much
lower than the final virial temperature of the system in order to prevent
hydrodynamical effects from becoming important before the gas turns around and
passes through the shock.  Two different simulations were performed, one with
$K_0=0$ and one with $K_0=0.1$. The simulations are evolved until the turnaround
radius encompasses half of the total number of particles. At the final time,
about $25\%$ of the particles have passed through the accretion shock.

Figures 4 and 5 show the dimensionless profiles of density, velocity, and
temperature averaged in spherical bins of constant logarithmic width. The
dimensionless mass enclosed inside each bin is also shown in the bottom right
panel. Each panel shows the result of the simulations at three different times,
corresponding to different numbers of particles within the shock radius: $1,838$
(triangles), $3,648$ (squares), and $5,443$ (circles) for the simulation without
cooling and $1,655$ (triangles), $3,371$ (squares), and $4,734$ (circles) for
the simulation with cooling. The analytic eigensolutions corresponding to
$K_0=0$ and $K_0=0.1$ are shown with a solid and dashed line, respectively.

The simulation without cooling is in all respects similar to that reported by
Navarro \& White (1993), except for the fact that we use about ten times more
particles. As shown in Figure 4, the density and velocity profiles agree
remarkably well with the analytic solution inside the shock radius
$\lambda_s$. As expected, agreement with the analytic solution improves as more
and more particles pass through the shock and the post-shock flow becomes better
resolved. This is especially noticeable in the mass panel, where it is seen that
at early times (triangles) the simulation deviates from the analytic solution
but that the system converges to the right solution at later times. Once about
$5,000$ particles have gone through the shock the agreement between the solution
and the experiment is remarkably good.  The convergence towards the similarity
solution is also convincingly demonstrated in the temperature panel, where the
linear temperature scale accentuates the discrepancies near the center.

One important conclusion from this analysis is that although numerical
limitations compromise the results of the numerical simulations at early times,
these discrepancies have no major effect on the behaviour of the system at late
times.  The main deviations from the analytic solutions actually happen beyond
the nominal shock radius. The shock is smoothed over several resolution lengths,
and even at late times only outside approximately $2 \lambda_s$ the preshock
flow solution is recovered in the simulations.

As discussed in \S2.1 and illustrated directly by the dashed and solid lines in
Figures 4 and 5, the similarity solution changes substantially when radiative
cooling is included. The shock radius moves inwards, the post-shock temperature
increases, the infall velocity is non zero all the way to the center, and the
density decreases at all radii because of the accumulation of cold material at
the center.

Figure 5 shows that the numerical results match very well these predicted
changes in the post-shock region. Inside $\lambda_s=0.1872$ the temperature,
mass, density, and velocity profiles are almost indistinguishable from the
similarity solutions. Remarkably, this is the case even when only $7\%$
($\approx 1,600$ particles) of the mass of the system has passed through the
shock, and the agreement is seen to improve as more and more particles go
through the shock. As noted above, the main shortcoming of the simulation
regards the width of the shocked region: particles are seen to respond to the
shock as far out as $\sim 3$ times the nominal shock radius. It is encouraging,
however, to note that the overall trend is correct, and that the volume of the
shocked region is substantially smaller than in the case where cooling is
neglected.

\section{Discussion}

One important goal of the comparison between simulations and similarity
solutions presented above is to assess the reliability of numerical techniques
currently being used to simulate the formation of galaxy-sized structures in the
universe. The present study indicates that, to a large degree, SPH methods give
results that are consistent with analytic solutions in test cases that involve
some of the major ingredients believed to play a significant role during galaxy
formation. In particular, our results show that SPH codes can reproduce
faithfully the changes in temperature, density, and velocity that are associated
with strong ``cooling flows'' onto a central perturbation. The mass inflow rates
are also accurately reproduced, and there is no indication that numerical
limitations lead to an undue increase or decrement in the amount of mass that
cools and flows to the central object.

This is important because it has been argued that SPH-like treatments of
numerical hydrodynamics may cause an artificial ``overcooling instability'' that
exaggerates the importance of radiative cooling losses and leads to the
formation of excessively massive gas concentrations at the center of dark halos
(Thacker et al 1998). Our study shows that this instability is not present in
the test cases we present here. The rate at which gas cools and gets accreted
onto a central clump is proportional to the mass accretion rate through the
shock radius and is consistent with that expected from the similarity
solution. 

There are, however, important differences between the cases considered in our
study and in that of Thacker et al. These authors consider gas cooling from a
hot gaseous halo {\it in hydrostatic equilibrium} within a dark halo and report
a strong dependence of the cooled mass on the numerical resolution of SPH
simulations. They also consider a more realistic cooling function that varies
with temperature and density in very different ways than the self-similar
cooling function we adopt here. Because of these caveats, it may be premature to
argue either for or against their findings on the basis of the simulations
presented here. It should be possible, however, to test their arguments using
the cooling wave similarity solutions derived by Bertschinger (1989). These are
a much closer analogue to the case considered by Thacker et al and direct
comparison between numerical experiments and Bertschinger's analytic solutions
should provide a definitive assessment regarding the effects of numerical
resolution on the cooling and accretion of gas at the center of dark halos. We
plan to carry out this comparison in the near future.

\section{Summary}

We have derived similarity solutions that describe the spherical collapse,
shock, and settling of collisional gas from scalefree perturbations in an
Einstein-de Sitter universe. Our study extends prior work on the subject by
taking into account the full effects of energy loss due to radiative cooling
processes, under the simplifying assumption that the cooling function is a
simple power-law of density and temperature. This choice ensures that the time
evolution of the system is self-similar by requiring that the cooling time of
the system at all times is a fixed fraction of the age of the universe. The
solutions take into account many of the processes that are relevant to the
assembly of the baryonic component of galaxies in a cosmological scenario:
gravitational infall, energy dissipation through shocks, pressure gradients,
radiative energy losses, and cooling flows. Analytic solutions such as the ones
outlined here are invaluable to assess the reliability and diagnose the
shortcomings of numerical techniques currently being used in cosmological
simulations. 

The tests we present here show that SPH simulations reproduce the analytic
solutions very well. No substantial deviations from the predicted central mass
accretion rates or from the temperature, density, and velocity profiles are
observed inside the shock radius when cooling is included. The region affected
by the shock is, however, is much larger than predicted: effects from the shock
are seen as far out as 2 or 3 times the nominal shock radius. Although this does
not seem to affect adversely the post-shock behaviour of the gas under the
simplifying conditions we adopt, it may have unwanted consequences in cases with
more complex infall geometry (such as mergers) or that involve a cooling
function with a more sensitive dependence on temperature and density than
assumed here. We hope that the work reported here will encourage further efforts
to test and improve the numerical treatment of the hydrodynamics of galaxy
formation.

\bigskip
\noindent 

This work has been partially supported by the National Science and Engineering
Research Council of Canada. RGB acknowledges the support of the PPARC rolling
grant ``Extragalactic Astronomy and Cosmology at Durham'' and the use of
STARLINK computing facilities. MGA acknowledges the hospitality of
the Physics Department of Durham University.

{}

\end{document}